\documentclass[12pt]{article}

\usepackage{amsmath, amsthm, amsfonts}
\usepackage{graphicx}
\usepackage{amssymb}
\usepackage{mathrsfs}
\usepackage{tikz} 
\usepackage{hyperref,cite}
\hypersetup{colorlinks=true,linkcolor= blue,citecolor=blue,urlcolor=blue}
\usepackage[usenames,dvipsnames]{pstricks}

\usepackage{parskip}
\numberwithin{equation}{section}

\usetikzlibrary{arrows,snakes}
\newcommand{\be}{\begin{equation}}                                                                                          
\newcommand{\ee}{\end{equation}}                                               
                                           
\long\def\begincomment#1\endcomment{} 
\definecolor{lightblue}{rgb}{.1,.4,.5}
\definecolor{brown1}{rgb}{.64,.43,.138}

\title{Holographic Brownian motion at finite density}
\author{Pinaki Banerjee{\footnote {pinakib@imsc.res.in} } \\ \\
  \small Institute of Mathematical Sciences\\  
  \small CIT Campus, Taramani, Chennai-600113, India \\
  \small and \\
  \small Homi Bhabha National Institute\\
\small Training School Complex, Anushakti Nagar, Mumbai 400085, India}
\date{}

\begin{document}

\hspace{15cm} IMSC/2015/12/08

{\let\newpage\relax\maketitle}

\renewcommand{\baselinestretch}{1.15}\normalsize

\abstract{Brownian motion of a heavy charged particle at zero and small (but finite) temperature is studied  in presence of finite density. We are primarily interested in the dynamics at (near) zero temperature which is holographically described by motion of a fundamental string in an (near-)
extremal Reissner-Nordstr$\ddot{\text{o}}$m black hole. 
We analytically compute the functional form of retarded Green's function for small frequencies and extract the dissipative behavior at and near zero temperature.}

\newpage 

\tableofcontents

\section{\label{sec:intro}Introduction}
AdS/CFT or more generally gauge/gravity duality \cite{Maldacena:1997re,Gubser:1998bc,Witten:1998qj,Witten:1998zw} has been serving 
as a great weapon in a theoretician's armory to study strongly coupled systems analytically
for almost last two decades. Although for most of the cases its predictions are 
qualitative, there are instances (see for example the famous $\eta/s$ computation in \cite{Policastro:2001yc})
when it relates very formal theoretical frame work to real life experiments. Since its discovery 
this duality has glued many phenomena appearing in apparently different branches of physics together. 
Studying Brownian motion of a heavy particle using classical gravity technique is one such example
\cite{deBoer:2008gu,Son:2009vu} where holography relates a statistical system to a gravitational one. The dual gravity description involves a long fundamental string 
stretching from the boundary of the AdS space into the black hole horizon. Numerous 
works \cite{Herzog:2006gh, Banerjee:2013rca, Caceres:2010rm, Chakrabortty:2013kra,
Gubser:2006nz,CasalderreySolana:2007qw, Fischler:2014tka,
Giataganas:2013hwa, Giataganas:2013zaa, Roychowdhury:2015mta,Banerjee:2015fed} have been done
elaborating on different aspects of this set-up. \\

\begin{figure}
    \centering
\includegraphics[width=100mm,scale=1]{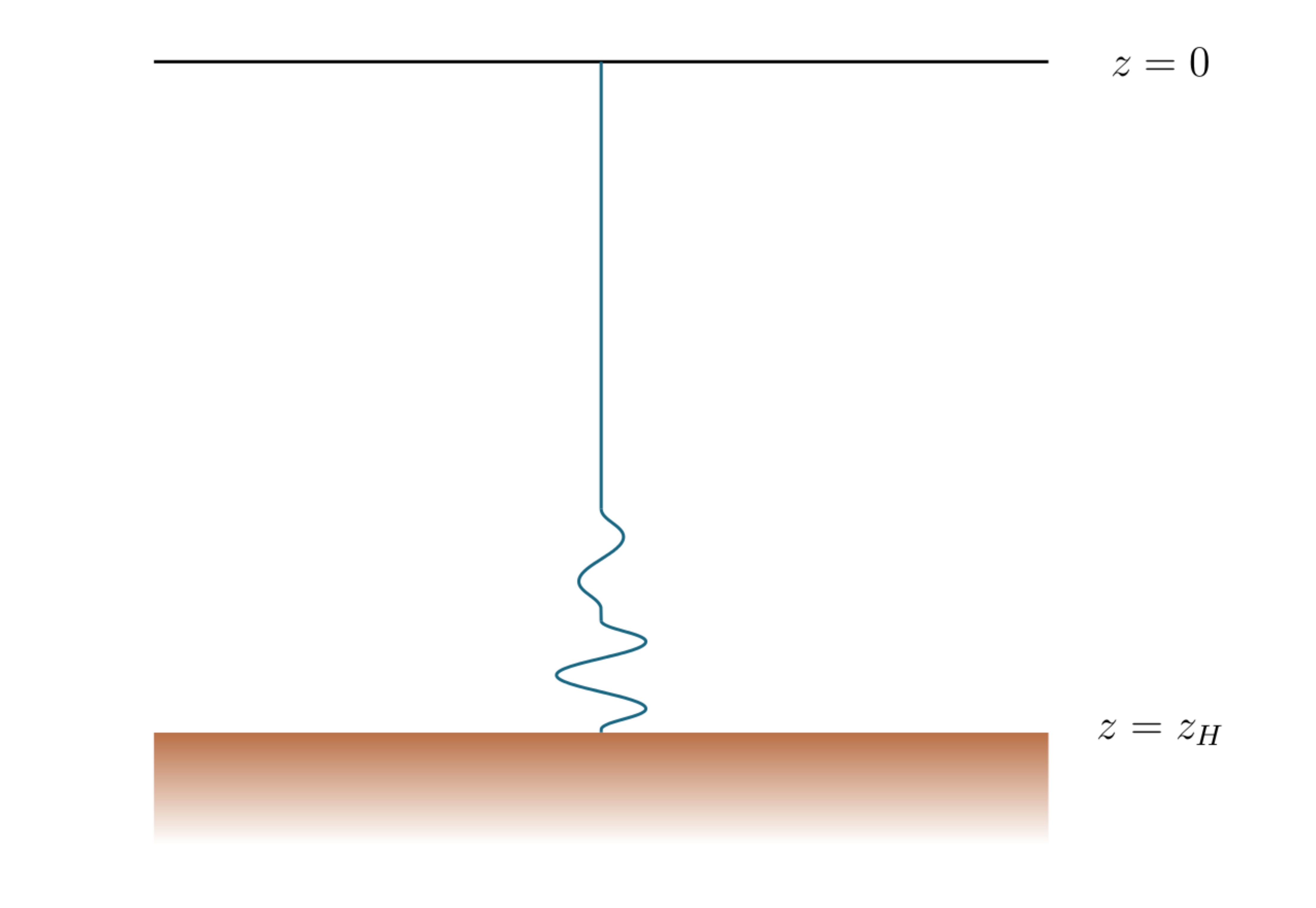}
    \caption{Gravity set up describing Brownian motion} \label{brownian}
\end{figure}

Integrating out\footnote{We mostly follow the Green's function language of \cite{Son:2009vu} to 
describe Brownian motion.} the
whole string in that background gives an effective description of the heavy particle at the 
boundary. Its dynamics is governed by a Langevin equation. For a particle with mass\footnote{We will see 
that this is actually `renormalized' mass. The correction to the \emph{bare mass}\,($M_0$) of the 
Brownian particle will come from the retarded Green's function.} $M$ which is
moving with velocity $v$ the Langevin equation reads
 
\begin{align}\label{eq:LE:1}
&M \frac{d v}{dt} +\gamma v =\xi(t) \\ \label{eq:LE:2}
\text{with \ \ \ }\langle &\xi(t)\xi(t^\prime)\rangle = \Gamma \delta(t-t^\prime) 
\end{align}

where $\gamma$ is the viscous drag, $\xi$ is the random force on the particle and $\Gamma$ 
quantifies the strength of the `noise' (i.e, random force). The Second equation is one of the 
many avatars of celebrated fluctuation-dissipation theorem. One can write down a 
generalized version\footnote{See, for example, \cite{Son:2009vu, Banerjee:2013rca} for a review of path integral
derivation of this generalized Langevin equation. Also notice that this equation is written
in terms of the \emph{bare mass} ($M_0$) of the Brownian particle.} of this equation 
\begin{align}\label{eq:LE:3}
M_0\frac{d^2x(t)}{dt^2}+ \int_{-\infty}^t dt^\prime ~G_R(t,t^\prime)x(t')=~\xi(t) \hspace{1.7cm} \langle\xi(t)\xi(t^\prime) \rangle =~ i G_{\text{sym}}(t,t^\prime)
\end{align}
$G_R(t,t')$ is thus the same as $\gamma(t-t')$ for the
choice of the lower limit of the integral, $t_0=-\infty$ and $iG_\text{sym}(t,t')$ is the same as $\Gamma(t-t')$. \\

In frequency space the generalized Langevin equation takes the following form
\begin{align}\label{eq:LE:4}
\left[- M_0 \, \omega^2 + G_R(\omega)\right] x(\omega)= \, \xi(\omega) \hspace{2cm} \langle\xi(-\omega) \, \xi(\omega) \rangle = \, i \, G_{\text{sym}}(\omega)
\end{align}

If the retarded Green's function, $G_R(\omega)$ is expanded for small frequencies 
the coefficient of $\omega^2 \left ( i.e, \frac{d^2x(t)}{dt^2} \right)$  adds to the bare mass 
of the particle and the coefficient of $\omega \left( i.e,\frac{dx(t)}{dt}\right)$ will
show off  as the drag term\footnote{More terms with higher powers in $\omega$ will also be generated
in this small frequency expansion. Their interpretations are outside the scope of standard
Langevin equation \eqref{eq:LE:1}. But from properties of Green's functions it is well known that
imaginary part of retarded Green's function causes dissipation. Thus odd powers in $\omega$ 
are responsible for dissipation. Actually the $\omega^3$ term signifies dissipation 
at zero temperature \cite{ Banerjee:2013rca, Banerjee:2015fed} in absence of matter density.}
\begin{align}\label{eq:LE:5}
G_R(\omega)=- i \gamma \omega -\Delta M \omega^2 + \ldots
\end{align}

After defining the `renormalized' mass as $$ M \equiv M_0 + \Delta M $$ this generalized Langevin
equations \eqref{eq:LE:4} (up to $\mathcal{O}(\omega^2)$) take the standard form \eqref{eq:LE:1} and \eqref{eq:LE:2}. \\

From the above discussion it is quite clear that if we are interested in studying dissipation 
for a Brownian particle we just need to compute the retarded Green's function, $G_R(\omega)$. We can
calculate this quantity using different holographic techniques \cite{Son:2002sd,Faulkner:2009wj} depending on the physical 
systems. \\ 

In \cite{Banerjee:2013rca} Brownian motion for a heavy quark in 1+1 dimensions was studied following
\cite{Son:2009vu} which used the prescription of \cite{Son:2002sd,Herzog:2002pc} to compute the boundary Green's function.
The calculation of \cite{Banerjee:2013rca} was performed in BTZ black hole background where 
the system is exactly solvable. The main result was to obtain dissipation for the heavy quark 
even at zero temperature. The result might look very counter intuitive and unphysical at 
first sight because at zero temperature the thermal fluctuations go to zero and therefore 
the Brownian particle should stop dissipating energy. But this zero temperature 
dissipation has its origin in radiation of an accelerated charged particle. The force term\footnote{ 
This force formula is same as ``Abraham-Lorentz force'' \cite{Griffiths} in classical electrodynamics only its ``coupling
constant'' (which is $\frac{q^2}{6 \pi \epsilon_0}$ for a particle with charge $q$ and $\epsilon_0$ is the vacuum permittivity) is different. This similarity is remarkable since in our holographic context the 
boundary theory is highly non-linear unlike electrodynamics!} in the
Langevin equation at zero temperature was of the form

\begin{align}\label{eq:LE:6}
F(\omega) =-i \, {\sqrt{\lambda}\over 2\pi} \, \omega^3 \, x(\omega)
\end{align}
Therefore the integrated energy loss is given by 
\begin{align}\label{eq:LE:7}
\Delta E = {\sqrt{\lambda}\over 2\pi}\int dt \, a^2
\end{align}

It is known from classical electrodynamics that the energy loss\footnote{This formula for an accelerated 
quark was obtained first by
Mikhailov in \cite{Mikhailov:2003er} by a very different approach. The factor $\sqrt{\lambda}\over 2\pi$ 
is essentially the Bremsstrahlung function 
$B(\lambda,N)$ ($2\pi B(\lambda,N)={\sqrt{\lambda}\over 2\pi} $) identified 
in \cite{Lewkowycz:2013laa} as occurring in many other physical quantities (such as the cusp 
anomalous dimension introduced by Polyakov \cite{Polyakov:1980ca}).} due to radiation is proportional to
square of the acceleration ($a$). See \cite{Chernicoff:2008sa,Chernicoff:2009re,Chernicoff:2009ff,Chernicoff:2011vn,
Chernicoff:2011xv,Xiao:2008nr,Hatta:2011gh,Caceres:2010rm,Fulton:1960,
Boulware:1979qj,Correa:2012at,Fiol:2012sg, Fiol:2015spa} for related works. \\

Dissipation at zero temperature is a fascinating phenomenon. Its emergence from the 
retarded Green's function signifies that $G_R(\omega)$ actually contains information at the 
`quantum' level (by `quantum' here we mean dynamics at $T=0$). Brownian motion of a particle is usually studied 
at finite temperature. The system is driven by fluctuations which are thermal in nature and
therefore if the temperature is taken to zero that $G_R(\omega)$ must vanish too. But the 
$G_R(\omega)$ we obtain from holography contains information of both thermal and quantum fluctuations
for the boundary theory. Although at finite $T$ thermal fluctuations dominate over the quantum
ones at very small $T$ the latter ones are much more important. The main aim 
of this paper is to understand how a heavy particle's (quark's) motion at finite density (chemical potential $\mu \ne 0$) 
is described at and near $T=0$. The dual gravity
theory should contain an (near-) extremal\footnote{The zero temperature dissipation for a theory dual to pure 
$AdS_{d+1}$ and black holes in $AdS_{d+1}$ has been calculated in \cite{Banerjee:2015fed}. Just on dimensional 
ground $G_R(\omega) \sim - i \omega^3$. The coefficient depends on the background. The cause of this 
dissipation being the radiation due to accelerated quark.} charged black hole. (See \cite{Edalati:2012tc,Ahmadvand:2015gfi} for some results 
in this set up).\\

For high temperature regime ($\mu \ll T$) the effect of the charge of the non-extremal black hole can be neglected
at the leading order and $G_R(\omega)$ can be computed in small $\mu$ and small $\omega$ expansions using the methods
followed in \cite{Son:2002sd,Son:2009vu}. \\
 
In this paper we would like to see how the system behaves near $T=0$. Therefore the other limit $\mu \gg T$ 
i.e, the low temperature regime is of more interest to us.
We will see that in this regime usual perturbation techniques for small $T$ and small $\omega$ won't work because
of a double pole for the $\omega^2$ term in the string equation of motion in the extremal black hole background.
Due to this double pole, near horizon dynamics is extremely  sensitive to $\omega$. To get around this problem we will 
adopt the  \emph{matching technique}\footnote{This matching technique is familiar to string theorists from the brane
absorption calculations that led to the discovery of AdS/CFT correspondence. For example see \cite{Klebanov:1997kc,Maldacena:1996ix}.
Maldacena used similar technique in his famous decoupling argument \cite{Maldacena:1997re}.}
described in \cite{Faulkner:2009wj} where the authors studied non-Fermi liquids using holography. \\

The rest of the paper is organized as follows. In section \ref{sec:Bckgd} we quickly review the
Reissner-Nordstr$\ddot{\text{o}}$m (RN) black hole in asymptotically AdS space time. The main purpose is to
spell out the notations and conventions that we will be following through out the article. The 
section \ref{sec:Zero} contains the analytic computation of retarded Green's function at zero temperature
using matching technique. We also list some of its properties in detail. The retarded Green's function at small but finite temperature is analyzed in section \ref{sec:Finite}. We mainly discuss how $G_R(\omega)$
gets small $T$ corrections. Section \ref{sec:Disc} summarizes the main results and their interpretations,
assumptions we make and some future directions. Section \ref{sec:Conc} contains some concluding remarks.

%

\section{\label{sec:Bckgd}AdS-RN black hole background}

AdS-RN black hole\footnote{The solution we will be working with has planer horizon with topology 
$\mathbb{R}^{d-1}$. Therefore it is really a black brane rather than a black hole.} is a solution to 
Einstein-Maxwell equation with a negative cosmological constant. 
\begin{align}
 S_{EM}= \frac{1}{2\kappa^2} \int d^{d+1}x \, \sqrt{-g} \, \left [ R + \frac{d(d-1)}{L_{d+1}^2} + \frac{L_{d+1}^2}{g_F^2} \, F_{MN} \, F^{MN} \right ]
\end{align}

$R$ is the Ricci scalar, $g_F$ is the dimensionless gauge coupling in the bulk. $L_{d+1}$ 
is a length scale (known as AdS radius)
and $\kappa^2$ is Newton's constant. Notice that we can always redefine the gauge field by absorbing 
the dimensionless coupling $g_F^2$ into $F_{MN}$. Thus we can fix $g_F$ to one 
without loss of generality. The (d+1) dimensional metric and gauge field that satisfy the corresponding equations of motion 
are given by
\begin{align}\label{AdS-RN}
ds^2 = \frac{L_{d+1}^2}{z^2}(-f(z)dt^2 + d\vec{x}^2) + \frac{L_{d+1}^2}{z^2} \frac{dz^2}{f(z)}
\end{align}
where, $$f(z)= 1+Q^2 z^{2d-2} - M z^d$$
$$A_t(z)= \mu (1-\frac{z^{d-2}}{z_0^{d-2}})$$ 
$Q,M,z_0$ are constant parameters which are black hole charge, black hole mass and horizon radius 
respectively. $\mu$ is the chemical potential,
$\vec{x} \equiv (‎x^1,x^2 \ldots x^{d-1})$ and 
$z$ is the radial co-ordinate in the bulk such that the boundary of this space is at $z=0$. \\

Notice that if we put $f(z)=1$ we get back pure $AdS_{d+1}$ in Poincare patch. This non trivial function 
$f(z)$ indicates that the physics changes as we move along the radial direction. \\ 

At the horizon : $f(z_0) = 0$. Therefore we can express $M$ as 
\begin{align}
 M = z_0^{-d} + Q^2 z_0^{d-2}
\end{align}

Now $Q$ can be expressed in terms of chemical potential($\mu$)
\begin{align}
 Q = \sqrt{\frac{2(d-2)}{d-1}} \frac{\mu}{z_0^{d-2}} 
\end{align}

And the Hawking temperature
\begin{align}
 T = \frac{d}{4\pi z_0} (1-\frac{d-2}{d} Q^2 z_0^{2d-2})
\end{align}

Actually $Q, M$ and $z_0$ are related to charge density, energy density and entropy density in the 
boundary theory respectively. $Q$ is charge density up to some numbers. Lets introduce a new length 
scale $z_*$ to express $Q$ as
\begin{align}
 Q := \sqrt{\frac{d}{d-2}}\frac{1}{z_*^{d-1}}
\end{align}
We also  define $\mu_* = \frac{1}{z_*}$. Note that $z_* \ge z_0$ to avoid the naked singular geometry.
(This is equivalent to $M \ge Q$ condition.) \\

There are two distinct situations possible : \emph{Extremal} ($T=0$) and \emph{Non-extremal} ($T \ne 0$). 

\subsection{Extremal black hole}

 When the Hawking temperature is zero the black hole is called extremal. Extremal black hole 
 contains maximum possible charge. The ``blackening function'' becomes 
\begin{align}\label{blackening-0}
 f(z) = 1+ \frac{d}{d-2} \frac{z^{2d-2}}{z_*^{2d-2}} - \frac{2(d-1)}{d-2} \frac{z^d}{z_*^d}
\end{align}

Near horizon region for the extremal black hole becomes $AdS_2 \times \mathbb{R}^{d-1}$

\begin{align} \label{AdS2}
 ds^2 &= \frac{L_2^2}{\zeta^2}(-dt^2 + d\zeta^2) + \mu_*^2 L_{d+1}^2 d\vec{x}^2 \\
 A_t(\zeta) &= \frac{1}{\sqrt{2d(d-1)}}\frac{1}{\zeta}
\end{align}
where, $\zeta := \frac{z_*^2}{d(d-1)(z_*-z)}$ , $L_2$ is the radius\footnote{Note that $L_2 < L_{d+1}$
for $d \ge 3$.} of the $AdS_2$ 
and is related to $L_{d+1}$ by the following relation
$$ L_2 = \frac{1}{\sqrt{d(d-1)}} L_{d+1} $$

\subsection{Non-extremal black hole}

Generically charged black holes have non-vanishing temperature. We will be interested in 
studying Brownian motion at finite density and finite temperature ($T$) but with $T \ll \mu$. 
We want to be in this regime because the near horizon geometry will become 
$AdS_2$-BH $\times \mathbb{R}^{d-1}$.

\begin{align} \label{AdS2-BH}
 ds^2 &= \frac{L_2^2}{\zeta^2}\left(-g(\zeta) dt^2 + \frac{d\zeta^2}{g(\zeta)}\right) + \mu_*^2 L_{d+1}^2 d\vec{x}^2 \\
 A_t(\zeta) &= \frac{1}{\sqrt{2d(d-1)}}\frac{1}{\zeta}(1-\frac{\zeta}{\zeta_0})
\end{align}
where $g(\zeta) := (1-\frac{\zeta^2}{\zeta_0^2})$,  $\zeta_0 := \frac{z_*^2}{d(d-1)(z_*-z_0)}$ and the corresponding temperature, 
$T = \frac{1}{2 \pi \zeta_0}$. For $T \approx \mu$ this nice structure breaks down.

\section{\label{sec:Zero}Brownian motion at zero temperature}
To understand Brownian motion of a heavy charged particle in some strongly coupled field theory in d-dimensions 
which has a gravity dual one needs to study the dynamics of a long string in the dual gravity
background \cite{deBoer:2008gu,Son:2009vu}. Therefore to explore the same Brownian motion at zero temperature 
and finite density one needs to study a string in an extremal charged black hole. This section contains the main analysis and results of the paper. \\

\subsection{Green's function by matching solutions}
In this Einstein-Maxwell theory an elementary string cannot couple to the gauge field, $A_M$. It can only couple
to the background metric $G_{M N}$. We consider geometries with vanishing Kalb-Ramond field, $B_{M N}$. For this
zero temperature case $G_{M N}$ can be read off from the extremal BH background \eqref{AdS-RN} with $f(z)$ given 
in \eqref{blackening-0}.\\

The string dynamics is given by standard Nambu-Goto action
\begin{align}
 S_{NG}= -\frac{1}{2 \pi l_s^2} \int d\tau \, d\sigma \, \sqrt{-h}
\end{align}

where $l_s$ is the string length and $h_{ab}$ is the induced metric on the world sheet
$$ h_{ab} = G_{MN} \, \partial_a X^M \, \partial_b X^N$$

We choose to work in static gauge, $$ \tau \equiv t  \text{\ \ and}~~ \sigma \equiv \zeta$$

Also  we can choose one particular direction, say $x_1$ (we call 
this simply $x$ for brevity), along which the world sheet fluctuates. 
$$ x \equiv x(\tau, \sigma) = x(t,\zeta)$$

To understand the dynamics of the string we need to use the full background metric \eqref{AdS-RN} with
the ``blackening factor'' given in \eqref{blackening-0}. Varying the Nambu-Goto action 
\begin{align}
 S_{NG} = -\frac{1}{2 \pi l_s^2} \int d t \, d z \, \frac{L_{d+1}}{z^2} \left[1 + \frac{1}{2} \, f(z)\, x'^2- \frac{1}{2f(z)} \, \dot{x}^2 \right]
\end{align}

we obtain the equation of motion (EOM) in frequency space  
 
\begin{align} \label{EOM-UV-0}
 x_\omega''(z) + \, \frac{\frac{d}{dz}(\frac{f(z)}{z^2})}{\frac{f(z)}{z^2}} \, x_\omega'(z) + \frac{\omega^2}{[f(z)]^2} \, x_\omega(z) = 0
\end{align}

where we have used $x(z, t) = \int \frac{d \omega}{2 \pi} \, e^{- \, i \omega t} x_\omega(z)$ \\

Now to obtain $G_R(\omega)$ the standard procedure would be to solve this equation and 
obtain it from the on shell action. But this procedure involves a few subtleties \cite{Faulkner:2009wj}. Firstly
this differential equation is not exactly solvable. Even if we are interested in $G_R(\omega)$
for very small frequencies ($\omega \ll \mu$) we cannot directly perform a perturbation expansion
in small $\omega$. Because at zero temperature the $f(z)$ has a double zero at the horizon (extremal
limit) and consequently $\omega^2$ term in the equation of motion generates a double pole at the 
horizon. Thus this singular term dominates at the horizon irrespective of however small $\omega$  
we choose. \\

To get around this difficulty we closely follow the matching technique in \cite{Faulkner:2009wj}. At first
we isolate the `singular' near horizon region from the original background. We already know that
the near horizon geometry is given by $AdS_2 \times \mathbb{R}^{d-1}$ \eqref{AdS2}. This is referred to as IR/inner 
region and the rest of the space time as UV/outer region. We can solve the string EOM exactly 
in this IR region and therefore the treatment will be non-perturbative in $\omega$. On the other hand
the $\omega$-dependence in UV region can be treated perturbatively as there is no more $\omega$-sensitivity.
The main task is to match the solutions over these two regions. The overlap between these to regions 
is near the boundary ($\zeta \to 0$) of the $AdS_2$ where 
$$\frac{1}{\mu} \ll \zeta \ll \frac{1}{\omega}$$ 
$$ \text{with} ~~~ z_*^2 \frac{\omega^2}{f(z)} \sim \omega^2 \zeta^2~~~ \text{is very \emph{small} }$$ 
$$ \text{and} ~~~ \mu \zeta \sim \frac{z_*}{z_*-z} ~~\text{remains \emph{large}}$$ 

The last two expressions ensure that the $\omega$ dependent term becomes small in EOM and we are still
near the horizon respectively. 

\subsection*{$\bullet~~$ Inner region}

For the string in $AdS_2 \times \mathbb{R}^{d-1}$ \eqref{AdS2}

\begin{align}
 \sqrt{-h} &= \frac{L_2^2}{\zeta^2} \, \sqrt{1 +  \frac{L_{d+5}^2 }{L_2^2} \, \mu_*^2 \, \zeta^2 (x'^2-\dot{x}^2)} \nonumber \\
 & \approx \frac{L_2^2}{\zeta^2} \left[1 + \frac{1}{2} d(d-1) \, \mu_*^2 \, \zeta^2 (x'^2-\dot{x}^2) \right]
\end{align}

The Nambu-Goto action 
\begin{align}
 S_{NG} = -\frac{L_2^2}{2 \pi l_s^2} \int d t \, d \zeta  \left[\frac{1}{\zeta^2} + \frac{1}{2} d(d-1)\, \mu_*^2 \, (x'^2-\dot{x}^2) \right]
\end{align}

Varying this action we get a very simple EOM for the string which is that of 
a free wave equation
\begin{align}
 x''-\ddot{x} = 0
\end{align}

To solve this linear EOM, the standard way is to go to the Fourier space

\begin{align}
 x(\zeta, t) = \int \frac{d \omega}{2 \pi} \, e^{-\, i \omega t} \, x_\omega(\zeta)
\end{align}

 The equation of motion reduces to 
 
\begin{align} 
 x_\omega''(\zeta)+\omega^2 \, x_\omega(\zeta) = 0
\end{align}

This is very well known differential equation with two independent solutions 
$$ x_\omega(\zeta) = e^{\pm i \omega \zeta}$$ 

As we are interested in calculating retarded Green's function we need to pick the one which
is \emph{ingoing} at the horizon ($\zeta \to \infty$). It's easy to see that $e^{+\, i \omega \zeta}$
is ingoing at the horizon. \\

Once we pick the right solution at the horizon we need to expand that near the boundary($\zeta = 0$) of 
the IR geometry i.e, $AdS_2 \times \mathbb{R}^{d-1}$
\begin{align} \label{Overlap-1}
 x_\omega(\zeta) = 1 + i \omega \zeta ~,~~~~~~~~ \text{near}~~  \zeta = 0
\end{align}

The ratio of normalizable to non-normalizable mode fixes the Green's function for the IR geometry
\begin{align}\label{IR-GF}
 \mathscr{G}_R(\omega) = i \omega 
\end{align}

(We will see in \eqref{boundary} that for string in $AdS_{d+1}$ non-normalizable and normalizable modes go as $z^0$ and $z^3$
respectively. Where as in $AdS_2 \times \mathbb{R}^{d-1}$ \eqref{Overlap-1} they go as $z^0$ and $z^1$ .)

\subsection*{$\bullet~~$ Outer region}

For the outer region  we need to solve the full EOM \eqref{EOM-UV-0}. But now as we are away
from the `dangerous' near horizon region we can do a small frequency expansion. At the leading order 
we can put $\omega = 0$. 
Let's say that the \eqref{EOM-UV-0} has two independent solutions $\eta^{(0)}_+$ and $\eta^{(0)}_-$ 
for $\omega = 0$. We can fix there behavior near the horizon ($z=z_*$) and near the boundary ($z=0$) 
by solving this equation near those regions.

\subsubsection*{Near horizon}
Near $z=z_*$ $$f(z) \approx d(d-1) \frac{(z_*-z)^2}{z_*^2}$$

The EOM reduces to 
\begin{align}
 x_\omega''(z) - \frac{2}{z_*-z} ~ x_\omega'(z) = 0
\end{align}

The two independent solutions are  $$c ~~\text{and}~~ \frac{z_*}{z_*-z}$$

Here $c$ is some constant which can be chosen to be 1. Since we need to match the inner and the outer
solutions near $z=z_*$ let's express these independent solutions in terms of $\zeta$. 
\begin{align}\label{Overlap-2}
 \eta^{(0)}_+ \to \left(\frac{\zeta}{z_*}\right)^0 ~~,~~ \eta^{(0)}_- \to \left(\frac{\zeta}{z_*}\right)^1 
\end{align}

\subsubsection*{Near boundary}

Near the boundary, $z=0$ we can approximate $f(z) \approx 1$ and consequently the EOM

\begin{align}
 x_\omega''(z) - \frac{2}{z} ~ x_\omega'(z) = 0
\end{align}

The solutions near  $z=0$ will behave as
\begin{align}\label{boundary}
 \eta^{(0)}_+ &\approx a^{(0)}_+\left(\frac{z}{z_*}\right)^0 + b^{(0)}_+  \left(\frac{z}{z_*}\right)^3 \\
 \eta^{(0)}_- &\approx a^{(0)}_-\left(\frac{z}{z_*}\right)^0 + b^{(0)}_-  \left(\frac{z}{z_*}\right)^3 
\end{align}

Notice that $a^{(0)}_{\pm} , b^{(0)}_{\pm}$ are not independent but related by Wronskian. 
We will use this information below to fix one of those coefficients.   

\subsubsection*{Matching the solutions}

We have some solutions to the full EOM in patches. All we need to do to obtain the Green's function 
is to determine the outer solution by matching it to the inner solution in the overlap region. Then expand that solution near $z=0$ to compute the ratio of normalizable to non-normalizable mode. \\

Let's do the matching first. From \eqref{Overlap-1} and \eqref{Overlap-2} we can express the
outer solution as 
\begin{align}\label{Outer-matched-1}
x_\omega(z) = \eta^{(0)}_+(z) + \mathscr{G}_R(\omega) z_* \eta^{(0)}_-(z)
\end{align}

Notice that so far we have been using solutions to the UV equation which are $0^{th}$-order in $\omega$
(as we have put $\omega= 0$). But in principle we can systematically add higher order corrections in 
$\omega$. In that improved version the outer solution will be given by

\begin{align}\label{Outer-matched-2}
x_\omega(z) &= \eta_+(z) + \mathscr{G}_R(\omega) z_* \eta_-(z) \\
\text{where}~~~ \eta_\pm(z) &= \eta^{(0)}_\pm(z) +\omega^2 \eta^{(2)}_\pm(z) + \omega^4 \eta^{(4)}_\pm(z) + \ldots 
\end{align}

And as before near boundary, $z=0$

\begin{align}\label{Outer-b'dy}
 \eta_\pm &\approx a_\pm\left(\frac{z}{z_*}\right)^0 + b_\pm  \left(\frac{z}{z_*}\right)^3 \\
 \text{where}~~~ a_\pm &= a^{(0)}_\pm +\omega^2 a^{(2)}_\pm + \omega^4 a^{(4)}_\pm + \ldots \\
 ~~~ b_\pm &= b^{(0)}_\pm +\omega^2 b^{(2)}_\pm + \omega^4 b^{(4)}_\pm + \ldots 
\end{align}

Note that $a_\pm,b_\pm$ are all \emph{real} coefficients because the UV equation \eqref{EOM-UV-0}
and the boundary condition \eqref{Overlap-2} at $z=z_*$ are both real. Also the perturbation in frequency are in even
powers in $\omega$ as \eqref{EOM-UV-0} contains only $\omega^2$. \\

Finally to obtain the retarded Green's function we expand the outer solution \eqref{Outer-matched-2}
near the boundary($z=0$) 
\begin{align}
 x_\omega(z) = A(\omega) \left(\frac{z}{z_*}\right)^0 + B(\omega) \left(\frac{z}{z_*}\right)^3
\end{align}

$$ A(\omega) = a_+ + \mathscr{G}_R(\omega) \, z_* \, a_- $$
$$ B(\omega) = b_+ + \mathscr{G}_R(\omega) \, z_* \, b_- $$

Green's function of the boundary theory is given by (see \cite{Son:2009vu,Banerjee:2013rca, Banerjee:2015fed})
\begin{align}
 G_R(\omega) := \lim_{z \to 0} T_0(z)\left(-\frac{z^2}{L_{d+1}^2}\right) \frac{{x_\omega}^\prime(z)}{x_\omega(z)}
\end{align}
where $$T_0(z) = \frac{1}{2 \pi l_s^2} \frac{L_{d+1}^4}{z^4} \left[1+ \frac{d}{d-2} \left(\frac{z}{z_*}\right)^{2d-2} -\frac{2(d-1)}{d-2} \left(\frac{z}{z_*}\right)^{d}\right],$$ 
is identified as local string tension which comes from the $z$-dependent normalization of the boundary
action. Since we are interested in boundary Green's function $$T_0(z) \approx \frac{1}{2 \pi l_s^2} \frac{L_{d+1}^4}{z^4}$$ 
and consequently the retarded Green's function

\begin{align}
 G_R(\omega) &\approx \lim_{z \to 0} \,- \frac{1}{2 \pi l_s^2} \, \frac{L_{d+1}^2}{z^2} \, \frac{{\eta_+}^\prime (z) + \mathscr{G}_R(\omega) z_* {\eta_-}^\prime(z)}{\eta_+(z) + \mathscr{G}_R(\omega) z_* \eta_-(z)} \nonumber \\
 &= \lim_{z \to 0} \, - \frac{L_{d+1}^2}{2 \pi l_s^2} \, \frac{1}{z^2} \, \frac{3 b_+ (\frac{z}{z_*})^2 \frac{1}{z_*} + \mathscr{G}_R(\omega) 3 b_- (\frac{z}{z_*})^2}{[a_+  + \mathscr{G}_R(\omega) z_* a_-] + [b_+  + \mathscr{G}_R(\omega) z_* b_-](\frac{z}{z_*})^3 } \nonumber \\ \label{G_R}
 &= - \frac{\sqrt{\lambda}}{2 \pi} \, \frac{3}{z_*^3} \, \left[ \frac{ b_+ + \mathscr{G}_R(\omega)  z_* b_-}{a_+  + \mathscr{G}_R(\omega) z_* a_-} \right]
\end{align}

We have introduced a dimensionless quantity $\lambda := \frac{L_{d+1}^4}{l_s^4}$ which behaves like a coupling constant 
in the boundary field theory. Since we are working in supergravity limit in the bulk $L_{d+1} \gg l_s$ and therefore $\lambda \gg 1$ i.e, 
the boundary theory is strongly coupled. The expression \eqref{G_R} is the main result of this paper. Below we analyze this in detail. \\

\subsection{Properties of the Green's function}

\bigskip
\begin{itemize}

 \item The expression \eqref{G_R} depends on two sets of data. 
 
 \begin{enumerate}
  \item \{$a_\pm, b_\pm$ \}  :  These constants come from solving EOM in the outer region.
  Therefore they depend on the geometry of the outer region. In this sense they are \emph{non-universal}
  UV data.
  
  \item $\mathscr{G}_R(\omega)$ : This depends only on the IR region which always contains $AdS_2$ 
  independent of the full UV theory. This is \emph{universal} IR data. 
 \end{enumerate} 
 
\bigskip

 \item As we have already pointed out the UV data ($a_\pm, b_\pm$) are always \emph{real}. Whereas 
 the IR data ($\mathscr{G}_R(\omega)$) is in general \emph{complex}. Therefore the dissipation is always 
 controlled by the IR data. Actually all non-analytic\footnote{There is no non-integer powers of $\omega$
 for the system we are considering. Therefore there is no branch cuts but
 $G_R(\omega)$ can only have poles at particular $\omega$-values.} behavior enters into $G_R(\omega)$ from 
 $\mathscr{G}_R(\omega)$. 
 
 \bigskip
 
 \item In principle $a^{(2n)}_\pm, b^{(2n)}_\pm$ are fixed by (numerically) solving the EOM in UV region in $\omega^2$
 perturbation.
 
 \bigskip 
 
 \item The interesting thing to notice is that the \eqref{EOM-UV-0} with $\omega=0$ allows a
 constant solution. From the boundary condition \eqref{Overlap-2} at $z=z_*$, we see that 
 $$\eta_+^{(0)}=1$$
 
 This value of $\eta_+^{(0)}$ will continue to solve the EOM \eqref{EOM-UV-0} with $\omega=0$
 for the outer region $z_* \ge z \ge 0$. So near the boundary ($z = 0$) we have (from \eqref{Outer-b'dy})  
 \begin{align}
   \eta_+ &\approx a^{(0)}_+\left(\frac{z}{z_*}\right)^0 + b^{(0)}_+ \left(\frac{z}{z_*}\right)^3 = 1
 \end{align}

This fixes $$ a^{(0)}_+ = 1 $$  $$ b^{(0)}_+ = 0 $$

 Actually we can fix one more coefficient by equating the generalized Wronskian\footnote{The generalized
 Wronskian of a 2nd order homogeneous ODE with two independent solutions $\phi_1$ and $\phi_2$ is defined as
 $$W(z) \equiv e^{\int^z P(t)dt} [\phi_1 \partial_z \phi_2 - \phi_2 \partial_z \phi_1]$$
 $$~~~~~~ = \sqrt{-g}g^{zz}[\phi_1 \partial_z \phi_2 - \phi_2 \partial_z \phi_1]$$} at the boundary
 and at the horizon. We get (see Appendix \ref{sec:B} for details) 
 $$ b_{-}^{(0)} = \frac{1}{3} $$
 
 The $0^{th}$-order Green's function reduces\footnote{There is no principle that 
 tells us that the all the coefficients of Green's function (even in $0^{th}$-th order in $\omega$) should be determined by analytic methods. Due to the simplicity of this particular differential equation we can fix few of them analytically. In general one needs numerical techniques to fix all of them. } to 
 \begin{align}\label{0-th-GF}
  G_R^{(0)}(\omega) &= - \frac{L_{d+1}^2}{2 \pi l_s^2} \, \frac{3}{z_*^3} \, \frac{ b^{(0)}_+ + \mathscr{G}_R(\omega) z_* b^{(0)}_-}{a^{(0)}_+ + \mathscr{G}_R(\omega) z_* a^{(0)}_-} \nonumber \\
  &= - \frac{\sqrt{\lambda}}{2 \pi} \, \frac{ i \mu_*^2 ~\omega}{(1 + i \frac{\omega}{\mu_*} \, a^{(0)}_-)}
 \end{align}

\begin{figure}
\centering
\includegraphics[width=100mm,scale=1]{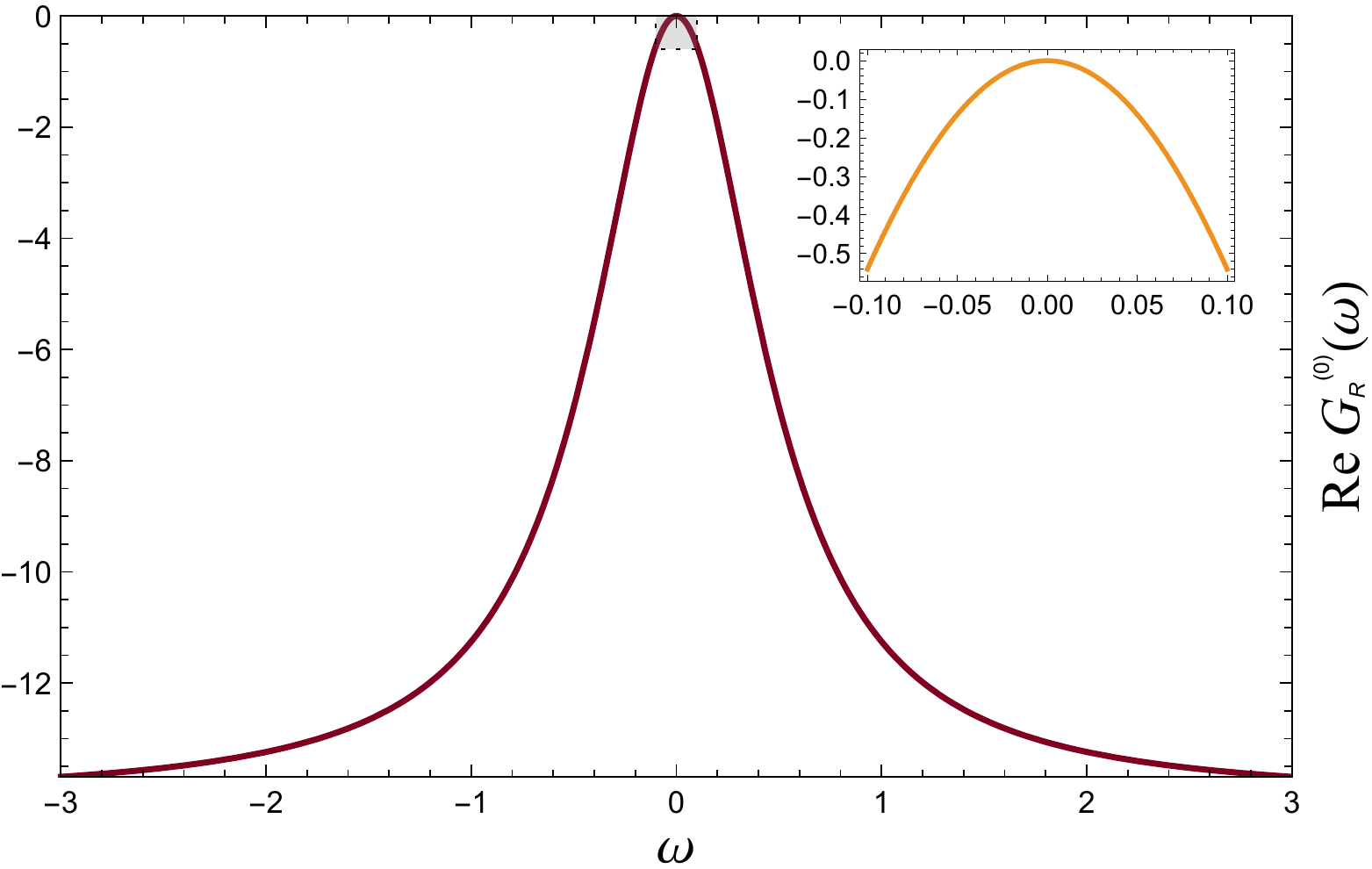}
\caption{Real part of  $G_R^{(0)}(\omega)$ with $\mu_* = 5$}\label{plot1}
\end{figure}

 The form of $G_R^{(0)}(\omega)$ ensures\footnote{Instead of a fluctuating string if one considers bulk Fermionic field (not world sheet field) in the same geometry,
 $a^{(n)}_\pm, b^{(n)}_\pm$ are functions of momentum $k$. For certain value of $k=k_F$ , say, $ a^{(0)}_+=0$.
 At this value of momentum $G_R^{(0)}(\omega,k_F)$ becomes singular at  $\omega = 0$. This indicates
 the \emph{Fermi surface}.} $$ G_R^{(0)}(\omega) =0 ~~~~~~~\text{as}~~  \omega \to 0 $$

The real and imaginary parts of $G_R^{(0)}(\omega)$ are plotted (see Fig.~\ref{plot1} and Fig.~\ref{plot2}) for particular values of the parameters : $\lambda=50, ~\mu_* = 5$ and $a^{(0)}_- = 10$. \\

For small frequency 
 \begin{align}\label{eq:small-freq}
 G_R^{(0)}(\omega) &= - i \frac{\sqrt{\lambda}}{2 \pi} \, \mu_*^2 \, \omega (1 - i \frac{\omega}{\mu_*} \, a^{(0)}_- + \ldots) \nonumber \\
 &\approx - i \, \frac{\sqrt{\lambda}}{2 \pi} \, \mu_*^2 ~\omega - a^{(0)}_-\, \frac{\sqrt{\lambda}}{2 \pi} \, \mu_* \,\omega^2
 \end{align}
 
\begin{figure}
\centering
\includegraphics[width=100mm,scale=1]{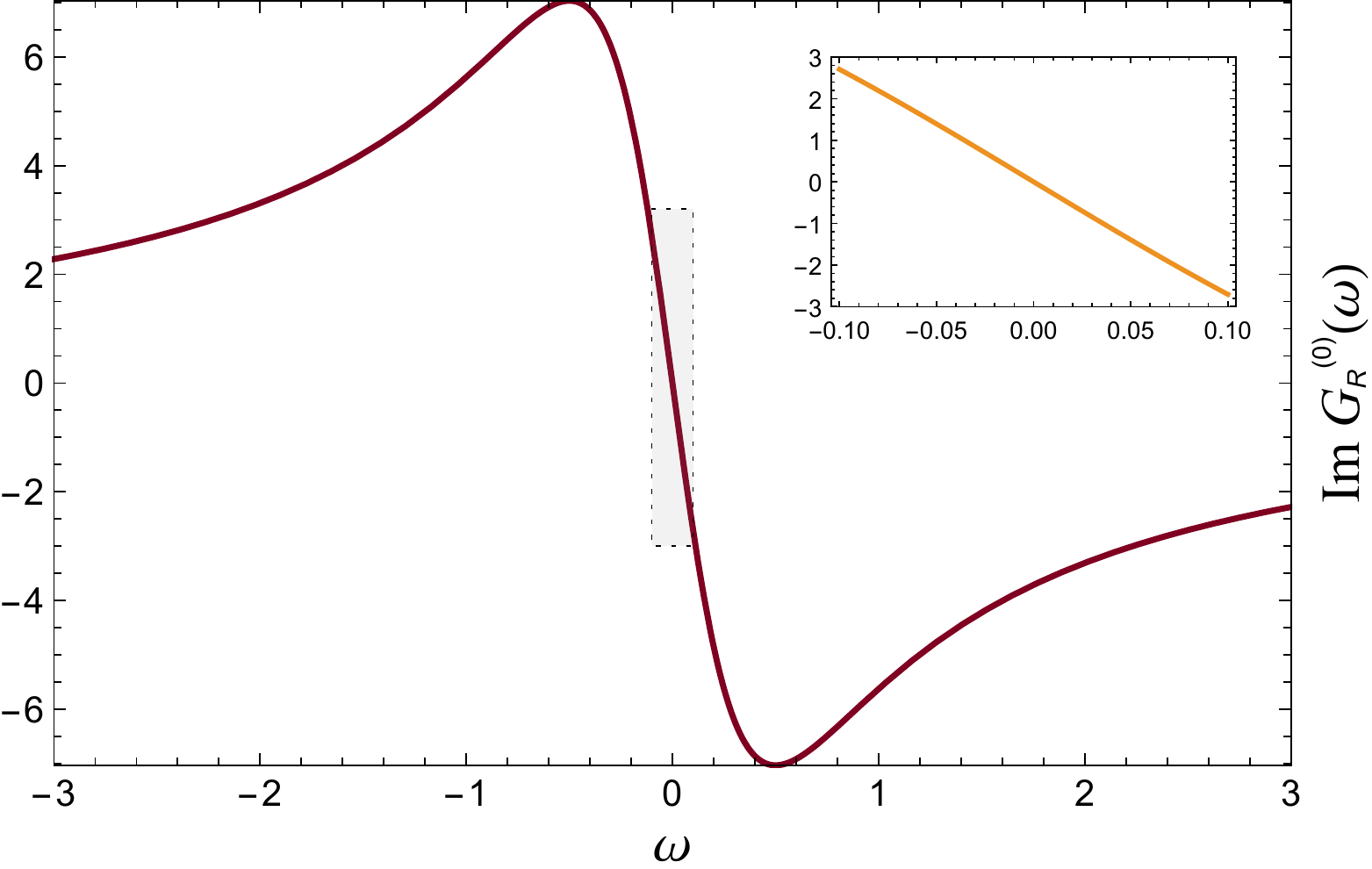}
\caption{Imaginary part of $G_R^{(0)}(\omega)$ with $\mu_* = 5$}\label{plot2}
\end{figure}

This is also consistent with Langevin equation \eqref{eq:LE:5} as $ G_R(\omega)$ expansion starts with $- i \omega$. Note that, for small frequencies, the zero temperature dissipation goes linear in $\omega$ (see Fig.~\ref{plot2}) unlike $\mu=0$ case \cite{Banerjee:2013rca,Banerjee:2015fed} where
this goes as $\omega^3$. The fact that $ G_R(\omega)$ is linear in $\omega$ comes from the fact that the effective $AdS_2$ dimension (which can be read off from \eqref{Overlap-2}) of the `quark operator' is \emph{one} (\emph{i.e.} $\Delta = 1$). \\

The leading dissipative term is proportional to $\mu_*^2$. This result indicates 
that energy loss for the charged Brownian particle is more for medium with higher charge density.

\end{itemize}

\section{\label{sec:Finite}Brownian motion at finite temperature}

To study Brownian motion at finite but very small temperature we need to follow the same steps as before.
But now the inner region will become a non-extremal (rather near extremal) black hole \eqref{AdS-RN}
background. 

\subsection{Green's function by matching solutions }

In this section we will go through the same procedure of matching functional form of the solutions 
in inner and outer regions. There will be few modifications to the zero temperature $G_R(\omega)$. 

\subsubsection*{$\bullet~~$ Inner region}

The metric in this region is black hole\footnote{This black hole is related to $AdS_2$ geometry by a co-ordinate transformation
\cite{Spradlin:1999bn, Faulkner:2011tm}
(combined with a gauge transformation) that acts as a conformal transformation on the boundary of $AdS_2$. 
So correlators can be obtained directly from $AdS_2$ correlators via conformal transformation. } 
in $AdS_2 \times \mathbb{R}^{d-1}$ \eqref{AdS2-BH}. \\

The Nambu-Goto action 
\begin{align}
 S_{NG} = -\frac{L_2^2}{2 \pi l_s^2} \int d t \,d \zeta  \left[\frac{1}{\zeta^2} + \frac{1}{2} d(d-1)\,\mu_*^2\,\left(g(\zeta)x'^2-\frac{1}{g(\zeta)}\dot{x}^2\right) \right]
\end{align}

Varying this action we obtain the EOM in frequency space
\begin{align} 
 x_\omega''(\zeta) + \,\frac{2 \,\zeta}{\zeta^2 - \zeta_0^2}\, x_\omega'(\zeta) + \frac{\zeta_0^4 \, \omega^2}{(\zeta^2 - \zeta_0^2)^2} \,x_\omega(\zeta) = 0
\end{align}

This EOM can be solved exactly. The two independent solutions\footnote{Notice the same $\zeta_0 \,tanh^{-1}\left(\frac{\zeta}{\zeta_0}\right)$ factor
appears in the conformal transformation from $AdS_2$ to $AdS_2$-BH (see \cite{Faulkner:2011tm}).} 
are : $$ x_\omega(\zeta) = e^{\pm \,i \,\zeta_0 \, \omega \, \tanh^{-1}\left(\frac{\zeta}{\zeta_0}\right)}$$

Again we are interested in retarded Green's function so we pick the solution which is ingoing at 
the horizon ($\zeta = \zeta_0$)  $$ e^{+\,i \,\zeta_0 \,\omega \, \tanh^{-1}\left(\frac{\zeta}{\zeta_0}\right)}$$.

Once we have the ingoing solution we need to expand it near the boundary($\zeta = 0$) of near horizon geometry
\begin{align}
x^R_\omega(\zeta) = 1 + i \, \omega \,\zeta 
\end{align}

We can now read off the Green's function in IR region
\begin{align}
\mathscr{G}_{R,T}(\omega) = i \, \omega  
\end{align}

This is identical to the zero temperature case \eqref{IR-GF}. \\

We have discussed earlier the dissipative part of $G_R(\omega)$ comes solely from
the IR Green's function. For this particular problem $\mathscr{G}_{R,T}(\omega) = \mathscr{G}_R(\omega)=  i \omega$.
Therefore $T$-dependence can only creep in via the expansion coefficients ($a_\pm, b_\pm$).

\subsubsection*{$\bullet~~$ Outer region}
This outer region analysis will be almost identical to that of the zero temperature case. One just has to be careful about the coefficients ($a_\pm, b_\pm$) which are now \emph{temperature dependent}, in general. Therefore we can skip repeating the analysis and directly write down the Green's function at finite temperature following the zero temperature case (see section \ref{sec:Zero})

\begin{align}\label{G_R-T}
 G_{R,T}(\omega) &= - \frac{\sqrt{\lambda}}{2 \pi} \, \frac{3}{z_*^3} \, \left[ \frac{ b_+(\omega, T) + \mathscr{G}_{R,T}(\omega) \, z_* \, b_-(\omega,T)}{a_+(\omega,T) + \mathscr{G}_{R,T}(\omega) \, z_* \, a_-(\omega,T)} \right] \nonumber \\
 &= - \frac{\sqrt{\lambda}}{2 \pi} \, \frac{3}{z_*^3} \, \left[ \frac{ b_+(\omega, T) + i \, \omega \,z_* ~ b_-(\omega,T)}{a_+(\omega,T) + i \, \omega \, z_* \, a_-(\omega,T)} \right]
\end{align}

If we consider only the leading order in $\omega$ (i.e, putting $\omega=0$ in the EOM), even for the non-extremal case, 
$x_\omega =  const.$ is again a solution. As before we can normalize it to one. By the same argument as in zero
temperature case 

$$ a^{(0)}_+ = 1 $$  $$ b^{(0)}_+ = 0 $$

Therefore the leading order Green's function is identical to that of zero temperature case \eqref{0-th-GF} 

 \begin{align}\label{0-th-GF_T}
  G_{R,T}^{(0)}(\omega) =  - \frac{\sqrt{\lambda}}{2 \pi} \, \frac{ i \mu_*^2 ~\omega}{(1 + i \frac{\omega}{\mu_*} \, a^{(0)}_-)}
 \end{align}

 This Green's function can be improved by solving the \eqref{EOM-UV-0} 
 perturbatively in $\omega$ and $T$. Actually the corrections will be in powers of $\frac{\omega}{\mu_*}$  
 and $\frac{T}{\mu_*}$. The corresponding real coefficients can also be obtained numerically in a systematic 
 fashion. 

\section{\label{sec:Disc}Discussions}

We have studied in detail the important properties of the retarded Green's function we 
obtained from the matching technique. It has a nice structure in terms of frequency 
(and also in temperature). We discussed that the dissipative (in general non-analytic) part
of the system is determined by the near horizon behavior i.e, the IR data of the system. On the other
hand the near boundary behavior i.e the UV data is always some analytic expansion in nature.
Actually these facts are compatible with our field theoretic  and geometric intuitions. \\

For generic  many body systems we know that IR physics can show non-analytic behavior but UV physics
can only give analytic corrections to that. From Renormalization Group (RG) point of view this matching technique can be thought of
as matching UV to IR physics at some intermediate energy scale fixed by the chemical potential ($\mu_*$). \\

Geometrically also this is expected. Dissipation is caused due to energy or `modes' disappearing 
into ``something''. In the bulk picture this can only happen near the horizon of a black hole where the modes fall into
the black hole and never come back. Whereas near boundary geometry is very
smooth and therefore no non-analytic behavior can be expected from that UV region. \\

It is worth mentioning that the leading order dissipative term at zero temperature is linear\footnote{This linear dependence in frequency comes from  the fact that effective AdS$_2$ dimension (see \eqref{Overlap-2}) of the `quark operator' is \emph{one} (\emph{i.e.} $\Delta = 1$) and is very crucial. Due to this particular low frequency behaviour the dissipative structure is qualitatively same at zero and finite temperature. If the dimension has been different from \emph{one}, the small $\omega$ expansion in \eqref{eq:small-freq} at zero temperature would have started with a different power (i.e. not linear) and the story would have been different from the $T \ne 0$ result.} 
in frequency unlike the zero density situations \cite{Banerjee:2015fed, Banerjee:2013rca} where it starts with cubic term ($\omega^3$).  Therefore this is actually the \emph{drag} term associated to the velocity of the charged particle rather than the acceleration of the same. A particle moving at constant velocity at zero temperature can dissipate energy for this set-up since the presence of finite charge density breaks Lorentz symmetry of the boundary theory explicitly. Nevertheless there will be dissipation due to acceleration of the charged particle as radiation at the subleading order. Expanding $G_R^{(0)}(\omega)$ \eqref{0-th-GF} in small frequency one can obtain the Bremsstrahlung function $B(\lambda)$ by collecting the coefficient of $\omega^3$. 
$$ B(\lambda) = \frac{\sqrt{\lambda }}{2 \pi } \ {(a^{(0)}_-)}^2$$ 

$a^{(0)}_-$ can be fixed by numerical technique. But this is obtained solving the string EOM \eqref{EOM-UV-0} only upto $\mathcal{O}(\omega^0)$. It will get corrections for higher orders in $\omega^2$ that can be taken into account systematically. \\

For a particular theory at \emph{finite density} but at \emph{zero temperature} if one can independently compute the Bremsstrahlung function, then that can be compared with the result obtained in our method. The standard and well known method of computing the Bremsstrahlung function is using supersymmetric localization technique (see e.g, \cite{Fiol:2012sg, Lewkowycz:2013laa, Fiol:2015spa}). But one would face following challenges\footnote{The author would like to thank Tomeu Fiol for pointing out this possibility and also the possible challenges.} to apply this technique at finite density.  Firstly one needs to, if possible, turn on background fields corresponding to finite density while preserving enough supersymmetry. Secondly, and more specific to the computation of the Bremsstrahlung function, finite density breaks conformal invariance. Some of the steps in computing the Bremsstrahlung function use explicitly conformal symmetry. Although the Bremsstrahlung function must exist for non-conformal theories, it may no longer be controlled by localization. \\

The method we have used to obtain the Green's function only required an $AdS_2$ factor near the horizon.
Therefore it should work even if the UV theory is non-conformal (not asymptotically AdS) but the IR geometry
has a $AdS_2$ factor. 
For example, instead of D3 branes one can look at D2 or D4 brane geometries. They are 
non-conformal \cite{Itzhaki:1998dd}. If for some charge density they flow to some $AdS_2$ then this procedure 
can be applied. Also, by the same argument, it can possibly work for some rotating extremal black hole
backgrounds too. \\

All our results are valid for \emph{large} chemical potential and \emph{small} temperature. If one is 
interested in studying Brownian motion in the opposite regime this technique can not be 
used. The reason being for $\mu \sim T$ the `nice' inner region structure breaks down. In that case
the small $\mu$ corrections can be computed using same tecniques used in  \cite{Son:2009vu, Banerjee:2013rca} but for a charged black hole background in AdS with very small charge.


\section{\label{sec:Conc}Conclusions }
In this paper we have used holographic technique to study Langevin dynamics of a heavy particle moving 
at finite charge density. We have studied this by computing retarded Green's function via solution matching
technique. Here are the main results. \\

\begin{itemize}

 \item Analytic form of retarded Green's function for small frequencies has been obtained at zero temperature. \\
 
 \item The drag force at zero temperature shows up as the leading contribution at small frequencies.  \\
       
 \item It is also been sketched how the retarded Green's function gets corrections due to small (but finite) temperature. The leading dissipative part (drag) remains identical to that in the zero temperature case. \\
 
 \item The drag term grows quadratically  with the chemical potential i.e, loss in energy of the Brownian particle is more for medium with higher charge density. \\
 
 \item The leading contribution to the Bremsstrahlung function $B(\lambda)$ is obtained with an unknown co-efficient $a^{(0)}_-$ which can be fixed by numerical method. Its corrections in $\frac{\omega}{\mu_*}$  and $\frac{T}{\mu_*}$
 can be computed systematically. 
 \end{itemize}

\subsection*{Acknowledgments}  

PB is grateful to B. Sathiapalan for fruitful discussions, suggestions on the manuscript and encouragement. He acknowledges useful conversation with Roberto Emparan, Tomeu Fiol and Carlos Hoyos. The author also would like to thank Rusa Mandal for helping with Mathematica.

\appendix

\section{\label{sec:B}Fixing coefficient using generalized Wronskian}

A field $\psi(z)$ satisfying a second order linear homogeneous differential equation 
\begin{align} \label{B:DE}
 \psi''(z) + P(z) \psi'(z) + Q(z) \psi(z) = 0
\end{align}
where $P(z)$ and $Q(z)$ are real the generalized Wronskian is defined as
\begin{align}\label{B:W}
 W(\psi_1, \psi_2; z) &:=  e^{\int^z P(t)dt} [\psi_1 \partial_z \psi_2 - \psi_2 \partial_z \psi_1] \\
 &= \sqrt{-g}g^{zz}[\psi_1 \partial_z \psi_2 - \psi_2 \partial_z \psi_1]
\end{align}
 where $\psi_1$ and $\psi_2$ are two solutions of \eqref{B:DE}. The interesting fact about
 this $W(z)$ is it is independent of $z$  
 $$\partial_z W(\psi_1, \psi_2;z) = 0 .$$ 
 Therefore 
 we can write 
 $$W(\psi_1, \psi_2; z) \equiv  W(\psi_1, \psi_2)$$ 
 
 Equation \eqref{EOM-UV-0} is exactly of the form \eqref{B:DE}. We know how its two independent
 solutions behave at the horizon ($z=z_*$) and at the boundary ($z=0$).
 The generalized Wronskian 
 \begin{align}
 W(\psi_1, \psi_2) &= \left(\frac{L_{d+1}^2}{z^2}\right)^{\frac{d+1}{2}} \frac{f(z)}{L_{d+1}^2} z^2 (\psi_1 \partial_z \psi_2 - \psi_2 \partial_z \psi_1) \\
 &= \left(\frac{L_{d+1}}{z}\right)^{d-1} (\psi_1 \partial_z \psi_2 - \psi_2 \partial_z \psi_1)
 \end{align}
 is independent of $z$. Therefore 
\begin{align}\label{B:W-match}
 W(\psi_1, \psi_2)\bigg |_{z=0} = W(\psi_1, \psi_2)\bigg |_{z=z_*} 
\end{align}

Now for extremal case 
$$ f(z) = 1 + \frac{d}{d-2} \left (\frac{z}{z_*}\right )^2 -\frac{2d-2}{d-2} \left (\frac{z}{z_*}\right )^d $$ \\

\begin{itemize}
 \item $f(z)\big |_{z=0} = 1$ \\
 \item $f(z)\big|_{z=z_*} \approx d(d-1) \frac{(z_*-z)^2}{z_*^2}$
\end{itemize} 

\vspace{5mm}

\begin{align} \label{B:LHS}
 \text{LHS of \eqref{B:W-match} : } ~~~ W(\psi_1, \psi_2)\bigg |_{z=0} &= \frac{1}{z^2}(\eta_+^{(0)} \partial_z \eta_-^{(0)} - \eta_-^{(0)} \partial_z \eta_+^{(0)}) \nonumber \\
 &= \frac{3}{z_*^3} (a_+^{(0)} b_-^{(0)} - a_-^{(0)} b_+^{(0)} )
\end{align}

\vspace{5mm}

\begin{align} \label{B:RHS}
 \text{RHS of \eqref{B:W-match} : } ~~~ W(\psi_1, \psi_2)\bigg |_{z=z_*} &= \frac{d(d-1)(z_*-z)^2}{z^2 z_*^2}(\eta_+^{(0)} \partial_z \eta_-^{(0)} - \eta_-^{(0)} \partial_z \eta_+^{(0)}) \nonumber \\
 &= \frac{d(d-1)(z_*-z)^2}{z^2 z_*^2}\left [1.~ \partial_z \left (\frac{\zeta}{z_*}\right ) - \left (\frac{\zeta}{z_*}\right )  \partial_z (1) \right ] \nonumber \\
 &= \frac{1}{z_*^3}
 \end{align}

 Equating \eqref{B:LHS} and \eqref{B:RHS} $$ a_+^{(0)} b_-^{(0)} - a_-^{(0)} b_+^{(0)} = \frac{1}{3} $$
 
and substituting $b_+^{(0)}=0$ and $a_+^{(0)} =1$ we get 

$$ b_-^{(0)} = \frac{1}{3}  $$

\setlength{\parskip}{20pt}
\linespread{0.8}
\bibliographystyle{bibstyle}
\bibliography{chemical}

\end{document}